\documentstyle[preprint,aps]{revtex}

\begin{document}
\draft
\preprint{ }
\title {Rigorous upper bound for the persistent current in systems with
toroidal geometry}

\author{ G. Vignale}
\address{Department of Physics, Missouri University, Columbia, MO
65211}
\date{\today}
\maketitle

\begin{abstract}

It is shown that the absolute value of the persistent current in a system with
toroidal geometry is rigorously less than or equal to $e \hbar N /4 \pi m
r_0^2$, where $N$ is the number of electrons, and $r_0^{-2} = \langle
r_i^{-2}\rangle$  is the equilibrium average of the inverse of the square of
the distance of an electron from an axis threading the torus.  This result is
valid in  three and two dimensions for arbitrary interactions, impurity
potentials,  and magnetic fields.
\end{abstract}


\narrowtext

The phenomenon of persistent currents occurs when an electronic  system is
placed in a magnetic
field:  at thermal equilibrium, an electric current  flows without
dissipation of energy. This effect is usually  studied in systems which are
topologically equivalent
to a torus, for example a  metal ring, or a  hollow cylinder.  The interesting
quantity
is  the flow of current through a cross section of the torus.

The persistent current exhibits a variety of behaviors, depending on both the
magnetic field and the
geometric  parameters of the system.
A first example is that of a {\it thin} metal ring, i.e.,  a ring whose
thickness is much smaller than its radius.  Mesoscopic versions of this system
have received particular attention in the past few years. The experiments
\cite {Levy} were done in the Aharonov-Bohm configuration, in which a weak
magnetic flux threads the ring, without significantly affecting the electrons
orbits.   In this case, the main physical effect is the ``twisting" of the
boundary conditions on the electron wave function,  leading to a persistent
current which is a periodic function of the threading flux, with period
$\Phi_0 = hc/e$ \cite {Butt}. The typical  magnitude  of the current is $e
v/L$, where $v$ is a characteristic velocity of propagation of an electron,
and $L$ is the length of the ring.  The current vanishes if the ring is made
larger and larger ($L \to \infty$), so this effect is a purely mesoscopic one.

A different behavior is obtained in  a   two-dimensional
ring, with inner and outer radii $R_1$ and $R_2$ ($R_1<R_2$), in the presence a
strong perpendicular
magnetic field $B$ such that the magnetic length
$l=(\hbar c/e B)^{1/2} <<R_2-R_1$.  This model has been studied by several
authors \cite
{Halperin,Avishai}. The magnetic field induces  currents flowing in opposite
directions at the inner
and the outer edges of the ring.  If $R_1$ and $R_2$ are macroscopically
different, the edge currents
do not cancel each other exactly, and one is left with a net current that
fluctuates violently as a
function of of electron number when the Fermi level is in a gap between
two Landau levels \cite {Avishai}.  A typical value  of the order of  a
fraction of $e
\omega_c$ ($\omega_c = eB/mc$ is the cyclotron frequency) has been reported in
a numerical study
\cite {Avishai}.

 In the special limit $R_1 \to 0$, the ring becomes a ``punctured disk", and
then it  has been found \cite {Avishai2} (neglecting disorder and interactions)
that the net current
is quantized in integral multiples of  $e \omega_c /4 \pi$  when the chemical
potential is pinned to
one of the Landau levels in the bulk.

In view of the diversity exemplified above, it is  remarkable what we show in
this paper, that there exists a rigorous upper bound to the persistent current
of a system of {\it arbitrary size and shape}, provided  that it is
topologically equivalent to a torus.  We do not need to assume any symmetry,
and we do not put any constraints on the nature of the magnetic field, not
even that it be uniform.  In essence, our derivation of the upper bound, is a
sharpening of the  argument  presented by Bohm \cite {Bohm}, following a
suggestion by Bloch, to prove that  a macroscopic  one-dimensional  ring
cannot carry a finite circulating current, at thermal equilibrium. Assuming
that the upper bound is a good estimate of the maximum value of the persistent
current that can be reached in a given system with an appropriate
magnetic field,  this  result enables us to easily understand the large
difference in order of magnitude and geometric dependence of persistent
currents in, for example,  thin rings and punctured disks.  Also, the rigorous
upper bound can be useful as a test of the validity of approximate theories of
persistent currents.

Let us consider a system of electrons confined  within a body of toroidal
topology, such as the one shown in Fig. 1. No symmetry is assumed.  Let us
choose an axis - the  ``$z$" axis - which threads the body, but is otherwise
arbitrary. The question of the optimal choice of the $z$ axis will be
addressed later.  Each half-plane emerging from the $z$ axis at an angle
$\phi$ ($0<\phi<2\pi$) cuts a two-dimensional cross section $S_\phi$ in the
body (see Fig. 1) \cite {Footnote}.  The persistent current is defined as the
flux of the current density $\vec j (\vec r)$ through $S_\phi$.  By virtue of
the continuity equation  $\vec \nabla \cdot \vec j (\vec r)=0$, this flux is
independent of the choice of $\phi$.

We introduce the standard  cylindrical coordinates $\vec r_i =
(r_i,z_i,\phi_i)$
to characterize the position of the i-th electron in the body.  The
corresponding  momenta are $\vec p_i = -i \hbar (\partial /\partial
r_i, \partial /\partial
z_i, r_i^{-1} \partial /\partial
\phi_i)$.  The Hamiltonian of the system is

\begin{equation}
\hat H = {1 \over 2m} \sum_i \left \{ \left (\vec p_i+{e \over c} \vec A (\vec
r_i)\right)^2 + V(\vec r_i) \right \} + {e^2 \over 2} \sum_{i \neq j} {1 \over
\vert \vec r_i - \vec r_j \vert} , \label{eq1}
\end{equation}
where $V(\vec r)$ and $\vec A(\vec r)$ are {\it arbitrary} scalar and vector
potentials.  The exact eigenfunctions
$\psi_n(r_1,z_1,\phi_1;.....r_N,z_N,\phi_N)$ of $\hat H$, with eigenvalue
$E_n$,
are completely antisymmetric, and vanish very rapidly  (exponentially)
outside the boundaries of the body.  The average persistent current, at
thermodynamic equilibrium, at temperature $T = 1/k_B \beta$ is given by

\begin{equation}
I = {1 \over Z} \sum_n e^{-\beta E_n}  \langle \psi_n \vert
\int_{S(\phi)} \hat j_\phi(\vec r)drdz \vert \psi_n \rangle, \label{eq2}
\end{equation}
where $Z$ is the partition function, and
\begin{equation}
\hat j_\phi(\vec r) = -{ e \over 2m} \sum_i \{ \hat \Pi_{i,\phi} \delta (\vec r
- \vec r_i) +\delta (\vec r - \vec r_i) \hat \Pi_{i,\phi}\},
\label{eq3}
\end{equation}
is the azimuthal component of the current density operator.  The azimuthal
component
of the kinetic momentum operator is defined as
\begin{equation}
\hat \Pi_{i,\phi}  = - {i \hbar \over r_i} {\partial \over \partial \phi_i} +
{e
\over c} A_\phi(\vec r_i).  \label{eq4}  \end{equation}
As we have already remarked, the integral, in  Eq.~(\ref{eq2}) is independent
of
the angle $\phi$ characterizing the cross section $S(\phi)$ in the $(r,z)$
plane.  Using this fact, together with the definition  of the current density
operator, Eq.~(\ref{eq3}), it is easy to verify that
 \begin{equation}
I =  \langle \hat I \rangle, \label{eq5}
\end{equation}
where
\begin{equation}
\hat I =   -{ e \over 2 \pi m} \sum_i {\hat \Pi_{i,\phi} \over r_i},
\label{eq6}  \end{equation}
is the current operator, and $\langle ... \rangle$ denotes the usual thermal
equilibrium average, with Hamiltonian $\hat H$.

The upper bound to $I$ is derived as follows.  Consider a gauge tranformation
\begin{equation}
\hat U_l =  \exp {[il \sum_i \phi_i]} , \label{eq7} \end{equation}
where $l$ must be an integer in order to preserve the single-valuedness of the
wave functions.  The Hamiltonian is transformed to
\begin{equation}
\hat H' =  \hat U_l \hat H \hat U_l^{-1} = \hat H - {l h \over e} \hat I +
{\hbar^2 l^2 \over 2m} \sum_i {1 \over r_i^2}. \label{eq8} \end{equation}
Let $F$ and $F'$ denote the free energies associated with the Hamiltonians
$\hat
H$ and $\hat H'$ respectively.  They satisfy the well known inequality \cite
{Feynman}
\begin{equation}
F' \leq  F + \langle \hat H' - \hat H \rangle.
\label{eq9} \end{equation}
But, in this case, $F' = F$ because $\hat H$ and $\hat H'$ are related by a
unitary transformation.  Therefore, combining  Eqs.~(\ref{eq9}) and
{}~(\ref{eq8}), we obtain the inequality
\begin{equation}
{\hbar^2 l^2 \over 2m} \langle   \sum_i {1 \over r_i^2}
\rangle - {l h \over e} I \geq  0, \label{eq10} \end{equation}
which must be satisfied for an arbitrary value of the integer $l$.  Let us
define
\begin{equation}
{1 \over r_0^2} \equiv {1 \over N}  \langle \sum_i {1 \over r_i^2} \rangle,
 \label{eq11} \end{equation}
and
\begin{equation}
I_m \equiv   {e \hbar N \over 4 \pi m r_0^2}.
 \label{eq12} \end{equation}
Then  Eq.~(\ref{eq10}) takes the form
\begin{equation}
l^2 - l {I \over I_m} \geq 0.
 \label{eq13} \end{equation}
It is now easy to verify that, for $\vert I \vert > I_m$, Eq.~(\ref{eq13})  is
always violated either by $l=1$ or by $l=-1$.  On the other hand, for
 $\vert I \vert \leq I_m$, Eq.~(\ref{eq13}) is satisfied by all integral
values of $l$.  We conclude that
\begin{equation}
\vert I \vert \leq I_m,
 \label{eq14} \end{equation}
the result of this paper.

The above argument is also straightforwardly applicable to a strictly
two dimensional ring geometry.  The arbitrary axis threading the toroidal body
is
replaced by an arbitrary point within the hole, and the cylindrical
coordinates relative to that axis are replaced by planar polar coordinates
relative to this point.  The cross section becomes a  segment (or a
collection of segments) of straight line, and the $z$ coordinate is suppressed.
The final result is  still given by Eq.~(\ref{eq14}). Incidentally, the fact
that the threading axis does not ``touch" the toroidal body guarantees that
$1/r_0^2$ is a finite number, because the wavefunction vanishes exponentially
in the regions where $r_i \to 0$.

Equation~(\ref{eq14}) is still dependent on the arbitrary choice of the
reference axis.  We can completely remove this arbitrariness, by choosing, for
any given system, the axis that yields the {\it most stringent} inequality,
i.e.
the smallest value of $1/r_0^2$.  Thus, our original definition of $r_0$ in
Eq.~(\ref{eq11}) is
replaced by
\begin{equation}
{1 \over r_0^2} \equiv min \{{1 \over N}  \langle \sum_i {1 \over r_i^2}
\rangle \},
 \label{eq15} \end{equation}
where the minimum is calculated with respect to the set of all possible axes
threading the toroidal body.  Qualitatively, this is the axis which, on the
average, remains as far as possible from the body. Notice that $1/r_0^2 <
1/r_{min}^2$, where
$r_{min}$ is the minimum distance from the optimal axis to the body.  Since
this latter quantity is
independent of the magnetic field, Eq.~(\ref{eq14}) implies that the current as
a function of
magnetic field cannot increase monotonically, but must either oscillate or tend
to a constant (at fixed particle number).

It is worth noting that the upper bound on the persistent current can be
expressed in terms of a
moment the electronic density distribution:
\begin{equation}
 I_m = {e \hbar \over 4 \pi m}\int {n (\vec x) \over r^2} d \vec x,
 \label{eq12bis}
 \end{equation}
where $r$ is the distance of point  $\vec x$ from the coordinate axis.
Thus, the calculation of $I_m$ in specific cases requires only a
knowledge of the electronic density distribution - a relatively ``gross"
property of a many-electron system.  Furthermore, from the form of
eq.~(\ref{eq12bis}), it is clear that the value of $I_m$ is not sensitive to
fine details of the density distribution, especially in bodies of
macroscopic to mesoscopic size.  Essentially, $I_m$ is a geometric
parameter. In three dimensional macroscopic bodies, we expect that
approximating the density as a constant within the geometric boundary of
the body and zero outside, should lead to an excellent approximation for
$I_m$.   A  better estimate of the density
distribution can be obtained by solving the Thomas-Fermi or the Kohn-Sham
equations.

  Let us now consider a few
concrete applications of Eq. ~(\ref{eq14}).  Suppose that the body is a
 circular ring of radius $R$ and negligible thickness (``thin ring").  Then the
optimal axis coincides with the axis of the ring, and $r_0 \simeq R$.  The
upper bound to the current is in this case
\begin{equation}
I_m  =  N { e h \over 2 m L^2},
 \label{eq16} \end{equation}
where $L=2 \pi R$ is the circumference of the ring. This tends to $0$ for $L
\to \infty$, {\it so there is no current flowing in a macroscopic thin ring}.
It must be noted that Eq. ~(\ref{eq16}) is rigorous for a geometrically thin
ring, i.e., it incorporates the effect of electron-electron and impurity
scattering exactly.  The ring may still be three-dimensional as far as the
electronic wave function is concerned.  No assumption on the form of the
density distribution has been used.  In the special case of a non-disordered
one-channel ring,  containing non-interacting electrons with a one-dimensional
Fermi velocity $v_F$, the upper bound  reduces  to $ 2 e v_F/L$ (the factor $2$
arising from spin degeneracy).  This value is attained exactly at $T=0$, when
the ring is threaded by a Aharonov-Bohm flux $\Phi = hc/2e$ (odd $N$) or $\Phi
= 0^+$ (even $N$).

As a second example, consider a  two dimensional disk (outer radius $R_2$)
with a  central hole of radius $R_1$,  in a strong magnetic field,
such that $l<<R_2-R_1$.  Once again, the optimal axis coincides with the axis
of
the disk.  Since the electronic areal density $n_0=N/A$ ($A$ is
the area of the ring between the inner and the outer edge) is
essentially uniform in the bulk of the system, we can calculate $1/r_0^2$ by a
taking a uniform average of $1/r^2$, which yields the result $1/r_0^2 = 2 \pi
\ln {(R_2/R_1)} /A$.   Substituting this in Eq.~(\ref{eq12}), we
obtain
\begin{equation}
I_m  =   {e \hbar \over 2m} n_0 \ln
{({R_2 \over R_1})}.
 \label{eq17} \end{equation}
The behavior of the  upper bound
depends only on the aspect ratio $R_2/R_1$.  If $R_1 \sim R_2$ to an accuracy
much smaller
than $R_1$ or $R_2$, we simply recover the thin ring geometry, and the
persistent current  vanishes.
In the limit $R_1 \to 0$ (or $R_2 \to \infty$), we obtain the ``punctured disk"
geometry.  The
upper bound diverges  logarithmically, allowing a large persistent current.
(The logarithmic divergence is of course cut off when $R_1$ becomes
comparable to the size of the edge region, in which the approximation of
constant density is no longer valid.)  In all other cases, the logarithmic
factor is a constant of order $1$, and the upper bound to the current can be
written as $I_m \sim e \hbar /2 \pi m a^2$, where $a$ is the average distance
between the electrons.

In the interesting case that the filling factor $\nu = 2 \pi l^2n_0$, rather
than the areal density, is kept constant in the bulk of the ring,
Eq.~(\ref{eq17}) takes the form
\begin {equation}
I_m  = {\nu e \omega_c \over 4 \pi} \ln {({R_2 \over
R_1})}.
\label{eq18}
\end{equation}
In this case the upper bound is proportional to
the intensity of the magnetic field, and the persistent current is allowed to
grow indefinitely with the latter ($N$ is now not constant).  A  persistent
current of the order of $e
\omega_c / 4 \pi$   (we assume  that the geometric parameters are fixed and
that $ln(R_2/R_1)$ is of order unity) is consistent with
the results of  quantitative calculations of the persistent current in an ideal
ring \cite {Avishai}.

These examples show that the rigorous upper limit derived in
this paper  can provide a reasonably good estimate of the
actual maximum value of the persistent current in a given toroidal system.  Of
course, since the sign of the current is left completely undetermined by our
arguments, we cannot say anything about the result of {\it averaging}, for
example, with respect to the impurity distribution in an ensemble of
mesoscopic rings.

We emphasize that the upper bound derived in this paper applies to the current
{\it in
thermodynamic equilibrium}. Our  argument is {\it not} applicable to the
current carried by a superconducting
ring, because, in that case, the current-carrying state is not the true
equilibrium state, but, rather,
a metastable state which cannot decay \cite {Schrieffer}.
This work has been supported by NSF Grant No. DMR-9100988. I thank Michael
Geller for useful discussions.

\newpage

\begin{figure}
\caption{Cylindrical coordinates for a toroidal body.}
\label{fig1}
\end{figure}


\begin{references}


\bibitem{Levy} L. P. L\'{e}vy et al., Phys. Rev. Lett. 64, 2074 (1990);
 V. Chandrasekhar et al., Phys. Rev. Lett. 67, 3578 (1991).


\bibitem{Butt} M. B\"{u}ttiker, Y. Imry, and R. Landauer, Phys. Lett. 96A, 365
(1983).


\bibitem {Halperin} B. I. Halperin, Phys. Rev. B 25, 2185 (1982).

\bibitem{Avishai} Y. Avishai, Y. Hatsugai, and M. Kohmoto, Phys. Rev. B 47,
9501 (1993).

\bibitem{Avishai2}  Y. Avishai and M. Kohmoto, Phys.Rev.Lett 71,
279 (1993).

\bibitem{Bohm} D. Bohm, Phys. Rev. 75, 502 (1949).

\bibitem{Footnote} The cross section $S_\phi$ is allowed to consist of several
disconnected pieces.

\bibitem {Feynman} R. P. Feynman, {\it Statistical Mechanics: A Set of
Lectures},  (Benjamin, 1979), Chapter 2.11.

\bibitem {Schrieffer} J. R. Schrieffer, {\it Theory of Superconductivity}
(Benjamin, 1971), Chapter
8.7.


\end{references}
\end{document}